\documentclass[12pt,a4paper]{article}
\usepackage{a4wide}
\usepackage{amsmath}
\usepackage{amssymb}
\usepackage{epsfig}
\usepackage{subfigure}
\usepackage{exscale}
\usepackage{float}
\usepackage{bbm}
\usepackage[numbers,sort&compress]{natbib}
\usepackage{pst-plot, pstricks,pst-math}
\usepackage{fancybox,amssymb,color}
\usepackage{graphicx}
\usepackage{pstricks, color, graphicx, epsfig, psfrag}
\usepackage{amsfonts,amsmath,amssymb,slashed}
\usepackage{dsfont}
\usepackage{bbm,bm}
\usepackage{fancyhdr, a4wide}
\usepackage[english]{babel}
\usepackage{subfigure}

\makeatletter
\@addtoreset{equation}{section}
\makeatother

\title{Partition Function in One, Two and Three Spatial Dimensions from Effective Lagrangian Field Theory}

\author{Christoph P.\ Hofmann$^a$ \\ \\
\normalsize {$^a$ Facultad de Ciencias, Universidad de Colima} \\
\vspace{0.3cm}
\normalsize {Bernal D\'iaz del Castillo 340, Colima C.P.\ 28045, Mexico} \\
\normalsize{e-mail: christoph.peter.hofmann@gmail.com}  \\}

\begin{document}

\maketitle

\begin{abstract} \normalsize

The systematic effective Lagrangian method was first formulated in the context of the strong interaction: chiral perturbation theory
(CHPT) is the effective theory of Quantum Chromodynamics (QCD). It was then pointed out that the method can be transferred to the
nonrelativistic domain -- in particular, to describe the low-energy properties of ferromagnets. Interestingly, whereas for
Lorentz-invariant systems the effective Lagrangian method fails in one spatial dimension ($d_s$=1), it perfectly works for nonrelativistic
systems in $d_s$=1. In the present brief review, we give an outline of the method and then focus on the partition function for
ferromagnetic spin chains, ferromagnetic films and ferromagnetic crystals up to three loops in the perturbative expansion -- an accuracy
never achieved by conventional condensed matter methods. We then compare ferromagnets in $d_s$=1,2,3 with the behavior of QCD at low
temperatures by considering the pressure and the order parameter. The two apparently very different systems (ferromagnets and QCD) are
related from a universal point of view based on the spontaneously broken symmetry. In either case, the low-energy dynamics is described by
an effective theory containing Goldstone bosons as basic degrees of freedom.

\end{abstract}


\maketitle

\section{Introduction}

While the methods used in particle physics tend to be rather different from the microscopic approaches taken by condensed matter
physicists, there is though one fully systematic analytic method that can be applied to both sectors. The effective Lagrangian method,
based on a symmetry analysis of the underlying theory, makes use of the fact that the low-energy dynamics is dominated by Goldstone bosons
which emerge from the spontaneously broken symmetry: chiral symmetry {SU(3)}$_R \times$ {SU(3)}$_L \to$ {SU(3)}$_V$ in Quantum
Chromodynamics (QCD), spin rotation symmetry O(3) $\to$ O(2) in the context of ferromagnets. The method thus connects systems as disparate
as QCD and ferromagnets from a universal point of view based on symmetry. The low-energy properties of the system are an immediate
consequence of the spontaneously broken symmetry, while the specific microscopic details only manifest themselves in the values of a few
effective constants. Still, as we are dealing with nonrelativistic kinematics in the case of the ferromagnet, apart from analogies, there
are important differences: most remarkably, the effective Lagrangian method, unlike for systems with relativistic kinematics, perfectly
works for ferromagnets in one spatial dimension ($d_s$=1).

While the low-temperature behavior of QCD was discussed more than two decades ago within effective field theory \citep{GL87,GL89}, the
low-temperature properties of ferromagnetic crystals, films and spin chains were considered only very recently within the effective
Lagrangian framework \citep{Hof02,Hof11,Hof12a,Hof12b,Hof13}. In the present article we review both QCD and ferromagnets, trying to build
a bridge between the particle physics and condensed matter communities. For example, so-called chiral logarithms, well-known in
Lorentz-invariant effective theories in $d$=3+1, also show up in the context of ferromagnets in $d$=2+1 dimensions (our notation is
$d$=$d_s$+1, where $d_s$ is the spatial dimension).

In the first part of this review, an outline of the effective Lagrangian method is provided, covering both relativistic and
nonrelativistic kinematics. In the second part we present the low-temperature expansions for the pressure and the order parameters, i.e.,
the quark condensate in QCD and the spontaneous magnetization in the context of ferromagnets. In particular, by considering the
suppression of loop diagrams and the absorption of ultraviolet divergences, we point out analogies and differences in the low-temperature
behavior of ferromagnetic systems and QCD.

As is well-known, QCD -- the theory of the strong interaction -- cannot be solved perturbatively at low energies where the QCD coupling
constant is not a small parameter. Therefore its effective theory - chiral perturbation theory (CHPT) -- based on an expansion in powers
of energy and momenta, rather than on an expansion in powers of the QCD coupling constant, represents an indispensable tool to explore the
low-energy domain of QCD. In the case of ferromagnets, the Heisenberg model can be solved at low energies by microscopic methods. Here,
effective Lagrangians thus represent an alternative scheme which is, however, more efficient than conventional methods such as spin-wave
theory.

A crucial point is that the effective Lagrangian technique is fully systematic and model independent, and does not resort to any
approximations or {\it ad hoc} assumptions. To appreciate the power of the effective method, we mention that -- until Dyson's monumental
work on the $d$=3+1 ferromagnet \citep{Dys56} -- it was unclear at which order in the low-temperature expansion of the spontaneous
magnetization the spin-wave interaction shows up. While the correct answer is $T^4$ \citep{Dys56,Zit65}, other researchers obtained
$T^{7/4}$, $T^{2}$ and $T^3$ (see Refs.~\citep{Kra36,Ope37,Sch54,Kra55,Man58,BH60,TH62a,SHEB63,Cal63,OH63}). Within the effective
Lagrangian framework, it was straightforward to confirm Dyson's result \citep{Hof02}. Likewise, the effective method allowed one to go
beyond Dyson, demonstrating that the first interaction correction to $T^4$ is of order $T^{9/2}$ \citep{Hof11}, whereas all other proposals
in the literature, $T^5$, $T^{13/2}$ and $T^{15/2}$, are incorrect \citep{MT65b,Cha01,Ach11}.

In $d_s$=1,2 the analogous question regarding the spin-wave interaction has largely been ignored in the many relevant articles
\citep{ML69,Col72,Tak71,KY72,Tak73,YK73,CL83,Lyk83,Sch85,TY85,YT86,Tak86,FU86,Sch86,LS87,Tak87a,Tak87b,CCL87,CAEK89,Kop89,Tak90,AA90a,AA90b,Yam90,DL90,Yab91,SSI94,NT94a,NT94b,NHT95a,NHT95b,RS95,TNS96,SSC97,HCHB01,KSK03,JIRK04,GPL05,DK06,SB06,GBT07,APPC08,JIBJ08,LCSWD11,DK12},
which are based on methods as diverse as spin-wave theory, Schwinger-boson mean-field theory, Green functions, scaling methods and
numerical simulations. The impact of the spin-wave interaction in ferromagnetic films and spin chains was first addressed systematically
and conclusively solved with effective Lagrangians in Refs.~\citep{Hof12a,Hof12b,Hof13}.

\section{Effective Lagrangian Field Theory}

The link between the underlying, or microscopic, theory -- QCD Lagrangian and Heisenberg Hamiltonian in the present context -- and the
effective theory is provided by symmetry. The effective action $\int \mbox{d}^dx {\cal L}_{\mbox{\scriptsize eff}}$ must be invariant under
all the symmetries of the underlying theory \citep{Wei79}. The essential point is that the terms appearing in the effective Lagrangian
${\cal L}_{\mbox{\scriptsize eff}}$ can be organized according to the number of space and time derivatives that act on the Goldstone boson
fields. At low energies, terms with few derivatives dominate the dynamics \citep{Wei79,GL84,GL85}.

We first consider Lorentz-invariant theories. In Quantum Chromodynamics, the underlying Lagrangian is given by
\begin{equation}
{\cal L}_{\mbox{\scriptsize{QCD}}} = - \mbox{$\frac{1}{2g^2}$} \mbox{tr}_c G_{\mu \nu} G^{\mu \nu} + {\bar q} i \gamma^{\mu} D_{\mu} q
- {\bar q} m q ,
\end{equation}
where $g$ is the strong coupling constant, $q(x)$ is the quark field, $G_{\mu \nu}$ is the field strength of the gluon field,
$m = \mbox{\normalsize diag}(m_u, m_d, m_s, \dots)$ is the quark mass matrix, and tr$_c$ denotes the trace of a color matrix.
In the chiral limit (i.e., when the quark masses $m_u, m_d, m_s$ are sent to zero) the above expression is invariant under the chiral
transformation {SU(3)}$_R \times$ {SU(3)}$_L$. The QCD vacuum, on the other hand, is only invariant under {SU(3)}$_V$, such that the
chiral symmetry is spontaneously broken. Goldstone's theorem \citep{Gol61} then implies that we have eight pseudoscalar mesons in the
low-energy spectrum of QCD, which are identified with the three pions, the four kaons and the $\eta$-particle. In the real world where the
quark masses are different from zero, the chiral symmetry of the QCD Lagrangian is not exact, but only approximate. Therefore these
particles are not strictly massless, but they represent the lightest degrees of freedom in the spectrum. Readers not familiar with QCD may
consult the pedagogic Ref.~\citep{Leu95} at this point. The derivative of the QCD Hamiltonian with respect to $m_q$ is the operator
$\bar{q}q$. The corresponding derivative of the free energy density $z$ thus represents the expectation value of $\bar{q} q$, i.e. the
quark condensate,
\begin{equation}
\langle \bar{q} q \rangle (T,m_q) = \frac{\partial z}{\partial m_q} .
\end{equation}

Spontaneous symmetry breaking is a prevalent phenomenon also in condensed matter physics, e.g. in ferromagnets, which, on the microscopic
level, are captured by the Heisenberg Hamiltonian augmented by the Zeeman term,
\begin{equation}
{\cal H} = -J \sum_{n.n.} {\vec S}_m \cdot {\vec S}_n - \mu \sum_n {\vec S}_n \cdot {\vec H} , \quad J = \mbox{\normalsize const} .
\end{equation}
The magnetic field $\vec H$ couples to the vector of the total spin. The summation only extends over nearest neighbors, and the exchange
coupling constant $J$ is purely isotropic. If the magnetic field is switched off, the Hamiltonian is symmetric under O(3) spin rotations.
The ground state of the ferromagnet ($J>0$), however, is invariant under O(2) only, such that the spin rotation symmetry is spontaneously
broken.

The magnetic field ${\vec H}$ hence plays a role analogous to the quark masses $m_q$: they are explicit symmetry breaking parameters.
Likewise, the magnetization,
\begin{equation}
\Sigma(T,H) = - \frac{\partial z}{\partial (\mu H)} ,
\end{equation}
is the analog of the quark condensate $\langle \bar{q} q \rangle (T,m_q)$. In particular, spontaneous magnetization (i.e., magnetization
in the limit ${\vec H} \! \to \! 0$) corresponds to a nonzero value of the quark condensate in the chiral limit $m_q \! \to \! 0$. Both
quantities are order parameters, their nonzero values signaling spontaneous symmetry breaking. Although the spontaneous symmetry breaking
pattern O(3) $\to$ O(2) gives rise to two magnon fields according to Goldstone's  theorem \citep{Gol61}, in a nonrelativistic setting only
one type of spin-wave excitation -- or one magnon particle -- exists in the low-energy spectrum of the ferromagnet
\citep{Lan66,GHK68,NC76}. Unlike in a Lorentz-invariant framework, there is no 1:1-correspondence between the number of Goldstone fields
and Goldstone particles.

After this brief review of the underlying theories (QCD Lagrangian and Heisenberg Hamiltonian) we now proceed with the discussion of the
corresponding effective theories. Chiral perturbation theory (CHPT) \citep{GL84,GL85} is well-established in particle physics. It exploits
the fact that the low-energy dynamics of QCD is dominated by pions, kaons and the $\eta$-particle, i.e. the Goldstone bosons of the
spontaneously broken chiral symmetry. These are the relevant degrees of freedom that appear in the effective Lagrangian
${\cal L}_{\mbox{\scriptsize eff}}$. The terms in ${\cal L}_{\mbox{\scriptsize eff}}$ are organized according to the number of space-time
derivatives: we are thus dealing with a derivative expansion, or equivalently, with an expansion in powers of momenta. The leading
effective Lagrangian is of momentum order $p^2$,
\begin{eqnarray}
{\cal L}_{\mbox{\scriptsize eff, QCD}}^2 & = & \mbox{$\frac{1}{4}$} {\cal F}^2 \mbox{tr} (\partial_\mu U \partial^\mu U^\dagger)
+ \mbox{$\frac{1}{2}$} {\cal F}^2 B \mbox{tr}\{ m (U+ U^\dagger)\} \nonumber \\
& & U = \exp (i \pi^a \lambda_a/{\cal F}) , \quad a = 1, \dots , 8 ,
\end{eqnarray}
where the matrix $U$ contains the eight Goldstone boson fields $\pi^a$, with $\lambda_a$ as Gell-Mann matrices. The structure of the above
terms is unambiguously fixed by chiral and Lorentz symmetry: these are the symmetries of the underlying theory which the effective theory
inherits. Note that, at this order, there are two {\it a priori} unknown low-energy constants, ${\cal F}$ and $B$, that are not determined
by the symmetries and hence have to be determined experimentally. Pseudoscalar mesons obey a relativistic dispersion law,
\begin{eqnarray}
\omega & = & \sqrt{c^2 {\vec k}^2 + c^4 {M}^2_{\pi}} , \quad M^2_{\pi} = M^2 + m_1 M^4 + m_2 M^6 + {\cal O}(M^8) \nonumber \\
& & M^2 = (m_u + m_d) B ,
\end{eqnarray}
where $M_{\pi}$ is the renormalized Goldstone boson mass, and $m_1,m_2$ involve low-energy (or effective) constants from the subleading
pieces ${\cal L}^4_{\mbox{\scriptsize eff, QCD}}$ and ${\cal L}^6_{\mbox{\scriptsize eff, QCD}}$ of the effective Lagrangian \citep{GL89}.

In Ref.~\citep{Leu94a}, the effective Lagrangian method was transferred to nonrelativistic systems. A crucial point of that analysis is
that for nonrelativistic kinematics, order parameters related to the generators of the spontaneously broken group, show up as effective
constants of a topological term which dominates the low-energy dynamics. This can not happen in a Lorentz-invariant setting
\citep{Leu94b}. For the ferromagnet, the leading contribution in ${\cal L}_{\mbox{\scriptsize eff, F}}$ is of momentum order $p^2$
\citep{Leu94a},
\begin{equation}
\label{leadingLagrangian}
{\cal L}^2_{\mbox{\scriptsize eff, F}}= \Sigma \frac{\epsilon_{ab} {\partial}_0 U^a U^b}{1+ U^3} 
+ \Sigma \mu H U^3 - \mbox{$\frac{1}{2}$} F^2 {\partial}_r U^i {\partial}_r U^i .
\end{equation}
The effective constant of the topological term, involving one time derivative (${\partial}_0$) only, is the (zero-temperature) spontaneous
magnetization $\Sigma$. The two real components of the magnon field, $U^a (a=1,2)$, are the first two components of the three-dimensional
magnetization unit vector $U^i = (U^a, U^3)$. Ferromagnetic magnons obey a quadratic dispersion law,
\begin{equation}
\label{DispRelFerro}
\omega(\vec k) = \gamma {\vec k}^2 + \mu H + \gamma_1 {\vec k}^4 + \gamma_2 {\vec k}^6 + {\cal O}({\vec k}^8) ,
\quad \gamma = \frac{F^2}{\Sigma} ,
\end{equation}
where the coefficients $\gamma_1, \gamma_2$ contain higher-order effective constants from ${\cal L}^4_{\mbox{\scriptsize eff, F}}$ and
${\cal L}^6_{\mbox{\scriptsize eff, F}}$ \citep{Hof02}. It is important to note that, unlike in CHPT, time and space derivatives are not on
the same footing in the case of nonrelativistic kinematics: two powers of momentum count as only one power of energy or temperature:
$k^2 \propto \omega, T$. Finally we point out that the leading-order effective Lagrangian ${\cal L}^2_{\mbox{\scriptsize eff, F}}$ is
space-rotation invariant, although the underlying Heisenberg model is not. Lattice anisotropies only start manifesting themselves in the
next-to-leading piece ${\cal L}^4_{\mbox{\scriptsize eff, F}}$ \citep{HN93}. Still, here we assume that ${\cal L}^4_{\mbox{\scriptsize eff, F}}$
and higher-order pieces in ${\cal L}_{\mbox{\scriptsize eff, F}}$ are space-rotation invariant -- this idealization does not affect our
conclusions.

\section{Ferromagnets and Quantum Chromodynamics} 

While the low-temperature properties of QCD have been derived a long time ago within effective Lagrangian field theory \citep{GL87,GL89},
the analogous systematic three-loop analysis of ferromagnets in $d_s$=1,2,3 was performed only recently in
Refs.~\citep{Hof11,Hof12b,Hof13}. What is most remarkable from a conceptual point of view is that the effective method perfectly works in
one spatial dimension in the case of nonrelativistic kinematics, whereas in a Lorentz-invariant setting the method fails in $d_s$=1. The
reason is that the perturbative evaluation of the partition function is based on the suppression of loop diagrams by some power of
momenta. This loop suppression depends both on the spatial dimension $d_s$ of the system and on the dispersion relation of its Goldstone
bosons \citep{Hof13}. For systems displaying a quadratic dispersion relation, a loop corresponds to the integral
\begin{equation}
\int \! d \omega d^{d_s} k \ {(\omega - \gamma {\vec k}^2)}^{-1} \ \propto \ p^{d_s} ,
\end{equation}
which, on dimensional grounds, is proportional to $d_s$ powers of momentum. In particular, loops related to ferromagnetic spin chains are
still suppressed by one power of momentum.

\begin{figure}
\includegraphics[width=14cm]{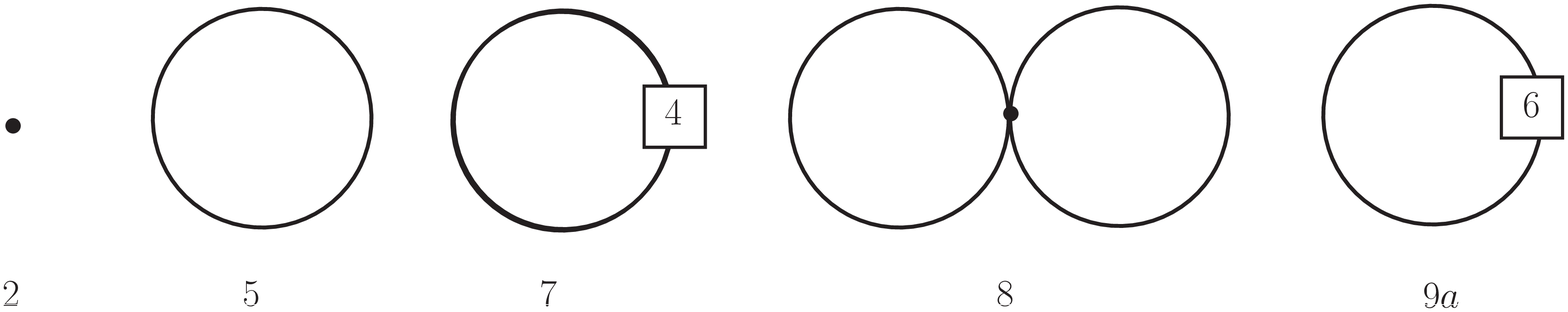}

\vspace{4mm}

\includegraphics[width=14cm]{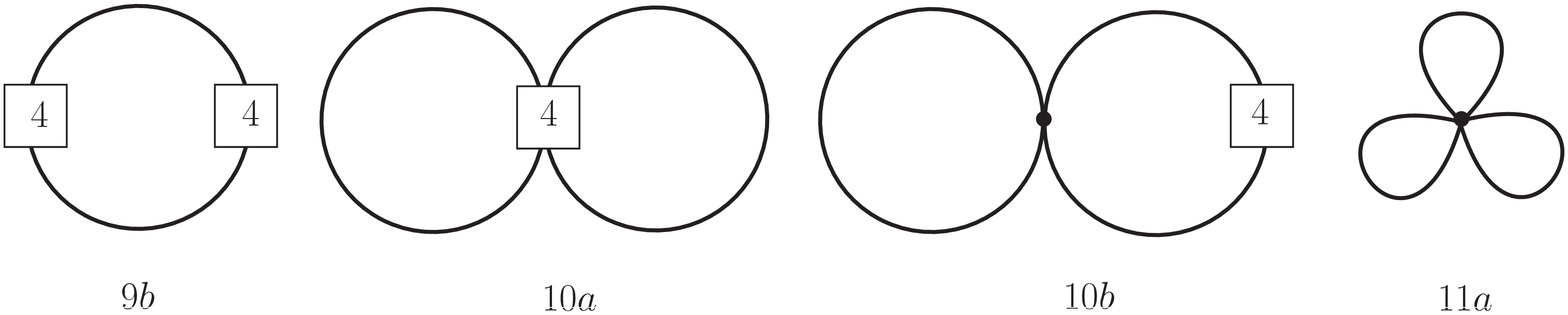}

\vspace{4mm}

\includegraphics[width=14cm]{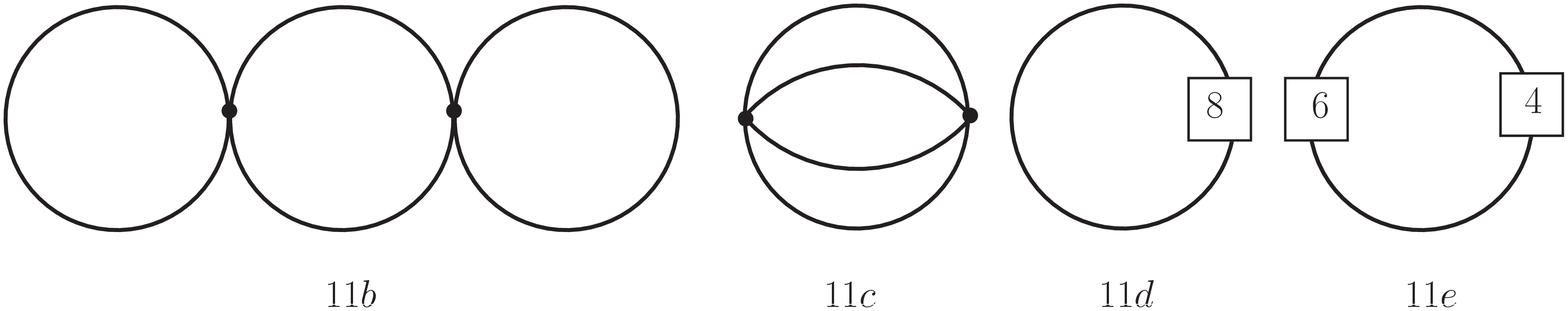}
\caption{Ferromagnet in three spatial dimensions: Feynman diagrams of the low-temperature expansion of the partition function up to three
loops ${\cal O}(p^{11})$. The vertices are numbered according to the piece of the effective Lagrangian ${\cal L}_{\mbox{\scriptsize eff}}$
they belong to. Vertices from the leading term ${\cal L}^2_{\mbox{\scriptsize eff}}$ are depicted by a filled circle. Loops in $d_s$=3 are
suppressed by three powers of momentum.}
\label{figure1}
\end{figure}

\begin{figure}
\includegraphics[width=14cm]{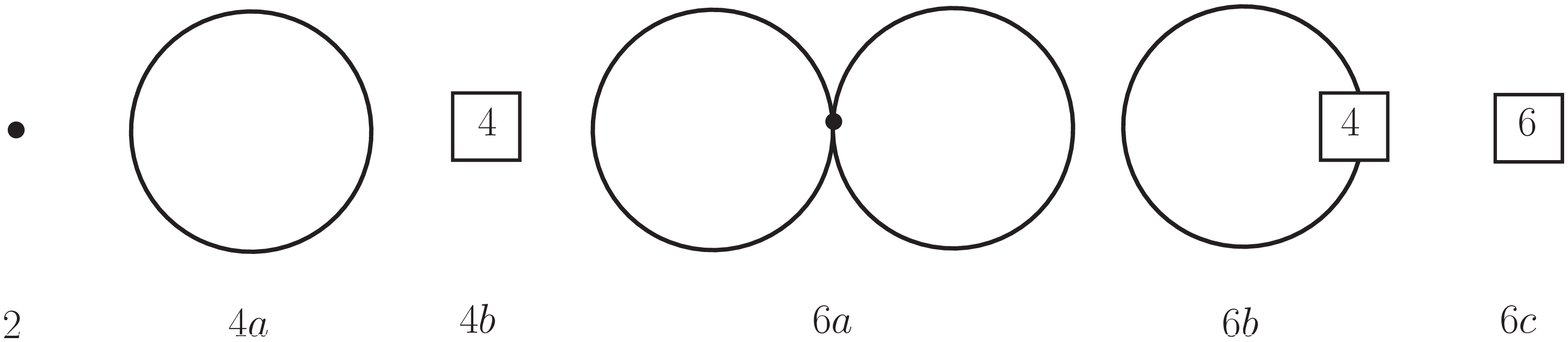}

\vspace{4mm}

\includegraphics[width=14cm]{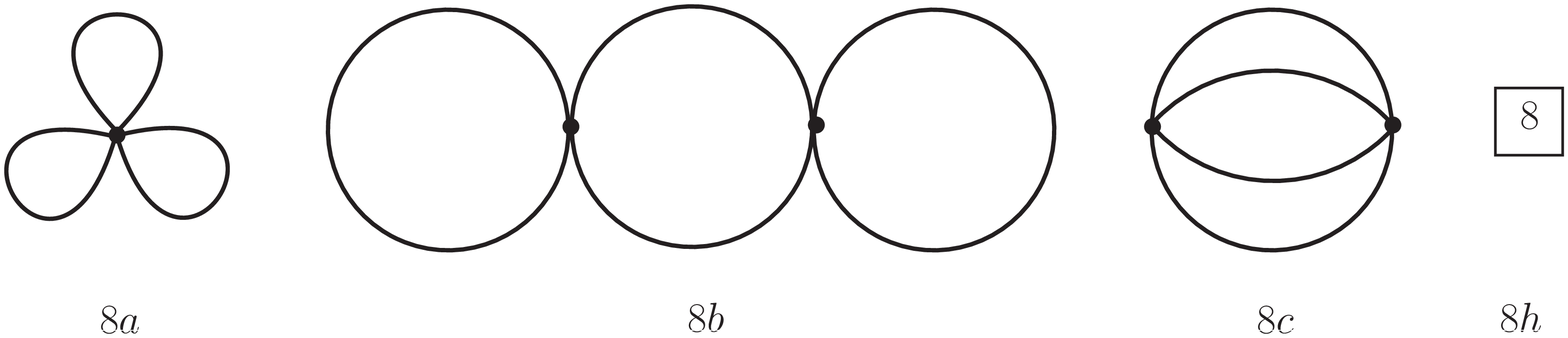}

\vspace{4mm}

\includegraphics[width=14cm]{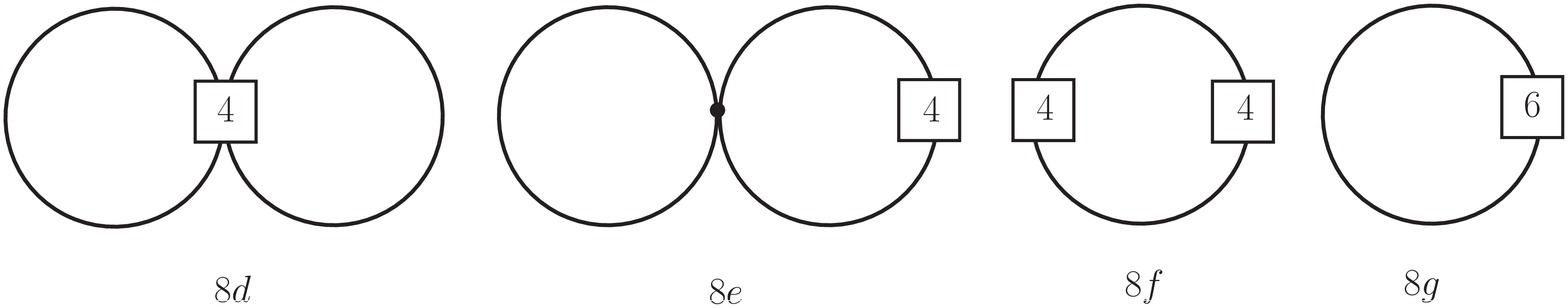}
\caption{Ferromagnet in two spatial dimensions and QCD: Feynman diagrams of the low-temperature expansion of the partition function up to
three loops ${\cal O}(p^8)$. Loops in $d_s$=2 are suppressed by two powers of momentum. In the case of the ferromagnet, the
temperature-independent diagrams 4b, 6c, 8h do not occur \citep{Hof02}.}
\label{figure2}
\end{figure}

\begin{figure}
\includegraphics[width=14cm]{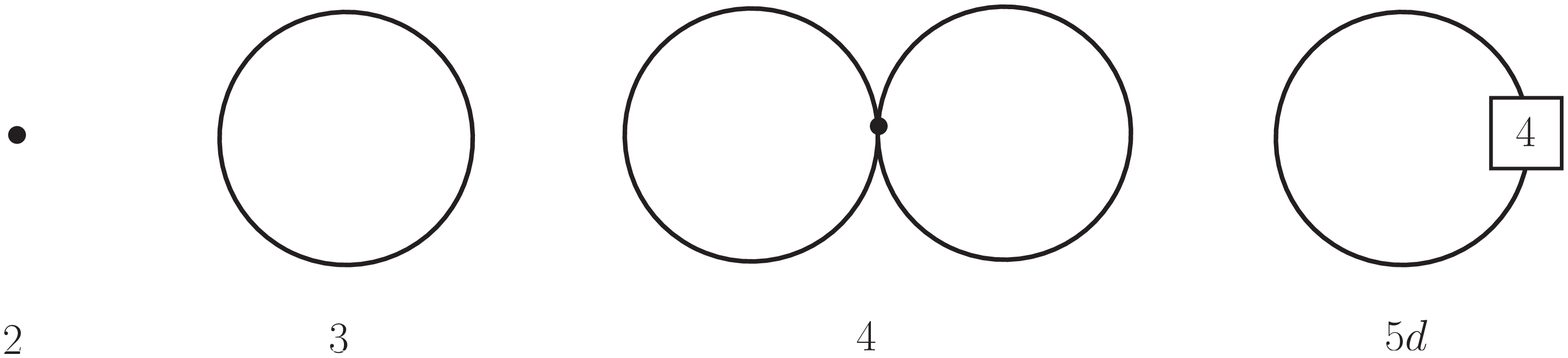}

\vspace{4mm}

\includegraphics[width=14cm]{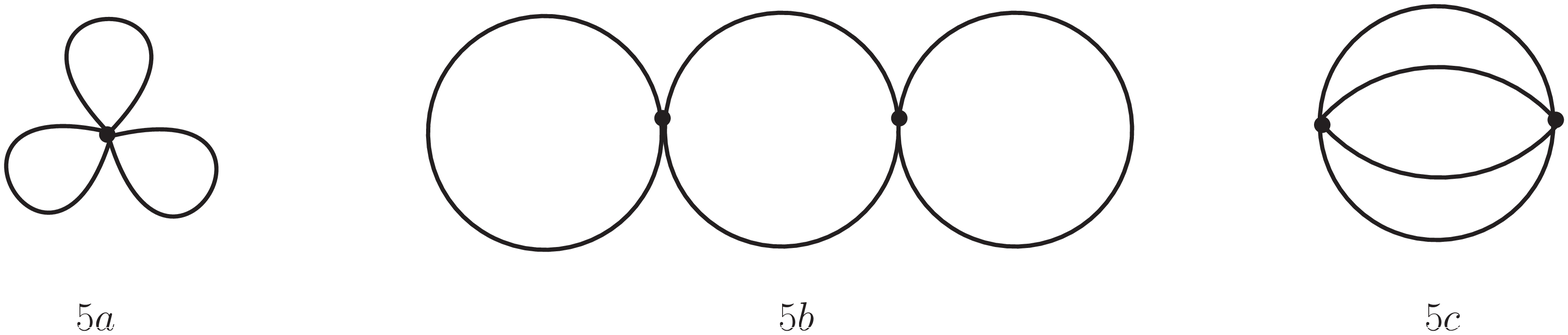}
\caption{Ferromagnet in one spatial dimension: Feynman diagrams of the low-temperature expansion of the partition function up to three
loops ${\cal O}(p^5)$. Loops in $d_s$=1 are suppressed by one momentum power.}
\label{figure3}
\end{figure}

For systems with a linear (relativistic) dispersion relation, obeyed e.g. by the pseudoscalar mesons in the chiral limit, the loop
suppression is rather different: there a loop involves the integral
\begin{equation}
\int \! d \omega d^{d_s} k \ {({\omega}^2 - c^2 {\vec k}^2)}^{-1} \ \propto \ p^{d_s-1} .
\end{equation}
For Lorentz-invariant systems, loops in $d_s$=3(2) are suppressed by two (one) power of momentum. However, in $d_s$=1, loops are not
suppressed at all, and the effective method fails to systematically analyze Lorentz-invariant systems in one spatial dimension. These
suppression rules are the basis for the organization of the Feynman graphs for the partition function of ferromagnets and QCD depicted in
Figs.~\ref{figure1}-\ref{figure3}.

The standard CHPT loop counting, depicted in Fig.~\ref{figure2}, corresponds to the loop counting for ferromagnets in two spatial
dimensions. Here each loop is suppressed by two powers of momentum. On general grounds, inspecting Figs.~\ref{figure1}-\ref{figure3}
reveals that in higher spatial dimensions, vertices involving subleading terms of the effective Lagrangian become more important. In
$d_s$=1, ${\cal L}^4_{\mbox{\scriptsize eff}}$ only appears in a one-loop diagram (diagram 5d, Fig.~\ref{figure3}), such that the
next-to-leading order effective constants in ${\cal L}^4_{\mbox{\scriptsize eff}}$ do not affect the spin-wave interaction. In $d_s$=2,3,
however, ${\cal L}^4_{\mbox{\scriptsize eff}}$ also shows up in two-loop graphs and thus does contribute to the spin-wave interaction. While
in $d_s$=2 these diagrams (8d and 8e) are of the same order as the three-loop graphs (diagrams 8a-c, Fig.~\ref{figure2}), in $d_s$=3
graphs 10a and 10b dominate over 11a-c. Furthermore, in $d_s$=2 we have insertions from ${\cal L}^6_{\mbox{\scriptsize eff}}$, and in $d_s$=3
even insertions from and ${\cal L}^8_{\mbox{\scriptsize eff}}$. In higher spatial dimensions, the symmetries thus become less restrictive:
more and more effective constants show up in a given thermodynamic quantity. Note however that the vertices which involve these
higher-order effective constants from ${\cal L}^6_{\mbox{\scriptsize eff}}$ and  ${\cal L}^8_{\mbox{\scriptsize eff}}$ only occur in one-loop
diagrams and hence do not contribute to the spin-wave interaction up to the three-loop level we are considering here.

Before we present the results, we have to discuss an important issue, well-known in chiral perturbation theory: graphs containing loops
are divergent in the ultraviolet and one has to take care of these singularities through renormalization. The basic object which creates
these singularities is the zero-temperature propagator at the origin: $\Delta(0)$. For Lorentz-invariant systems, the dimensionally
regularized expression is
\begin{equation}
\Delta(0) = {(4 \pi)}^{-d/2} \Gamma(1-d/2) M^{d-2} .
\end{equation}
In a relativistic setting, UV-divergences hence arise in even space-time dimensions $d$. In CHPT, these singularities are absorbed by
subleading effective constants in ${\cal L}_{\mbox{\scriptsize eff}}$, order by order in the derivative expansion in a systematic manner
\citep{GL84,Leu95,GL89}. In the case of nonrelativistic kinematics, the situation is rather different. Regarding ferromagnets, the
relevant expressions in the limit ${\vec x}, x_4 \! \to \! 0$ (where $x_4$ is Euclidean time) vanish in dimensional regularization. As a
consequence, only the temperature-dependent, i.e., finite, pieces are relevant, such that here the handling of "UV-divergences" is much
simpler \citep{Hof11,Hof12b,Hof13}.

The cateye diagram -- graph (11c,8c,5c) of Figs.~(\ref{figure1},\ref{figure2},\ref{figure3}) -- however, is more involved: as it does not
factorize into products of $\Delta(0)$ or derivatives thereof, the structure of the corresponding UV-divergences is more complicated. In
chiral perturbation theory, the ultraviolet divergences of graph 8c are absorbed by next-to-leading-order (NLO) effective constants from
${\cal L}^4_{\mbox{\scriptsize eff}}$ contained in the two-loop diagrams 8d and 8e. These NLO constants then undergo logarithmic
renormalization \cite{GL89}. For the ferromagnet in $d_s$ spatial dimensions, the cateye diagram is proportional to
\begin{equation}
\label{singular}
T^{d_s+2} \, {(\mu H)}^{\frac{d_s-2}{2}} { \Bigg\{ \sum_{n=1}^{\infty} \, \frac{e^{- \mu H n /T}}{n^{\frac{d_s+2}{2}}} \Bigg\} }^2
\, \Gamma(1-\frac{d_s}{2}) .
\end{equation}
In the case of ferromagnetic films ($d_s \! \to \! 2$), the above regularized expression is divergent. Much like in chiral perturbation
theory, the UV-singularity is absorbed by ${\cal L}^4_{\mbox{\scriptsize eff}}$-constants which are renormalized logarithmically. On the
other hand, in one and three spatial dimensions, the $\Gamma$-function does not develop a pole, and the next-to-leading order effective
constants do not undergo logarithmic renormalization -- they are finite as they stand.

The fact that there are no ultraviolet singularities in $d_s$=1,3 is crucial for the nonrelativistic effective framework to be consistent:
in $d_s$=1 (Fig.~\ref{figure3}) and $d_s$=3 (Fig.~\ref{figure1}), the cateye graph is of different order than the two-loop contributions
with vertices from ${\cal L}^4_{\mbox{\scriptsize eff}}$. In contrast to $d_s$=2, where three-loop and two-loop graphs "communicate" (they
are of the same momentum order $p^8$) and the UV-singularities can thus be absorbed by ${\cal L}^4_{\mbox{\scriptsize eff}}$ constants, the
effective loop analysis appears to be inconsistent in $d_s$=1,3: there are no two-loop diagrams available to absorb the "divergences" of
the cateye graphs 11c (Fig.~\ref{figure1}) and 5c (Fig.~\ref{figure3}). The puzzle is solved by noticing that in $d_s$=1,3, the cateye
graph is not divergent and that the perturbative scheme hence is perfectly consistent.

We end our discussion of partition function diagrams by pointing out that, on the two-loop level, there is an important difference between
chiral perturbation theory and ferromagnets. While graph 6a (Fig.~\ref{figure2}) in CHPT does contribute to the partition function, the
same diagram for ferromagnets in any $d_s$ (i.e., including graph 8 of Fig.~\ref{figure1} and graph 4 of Fig.~\ref{figure3}) turns out to
be zero because of parity \citep{Hof02,Hof12a,Hof13}. As a consequence, in the thermodynamic quantities the spin-wave interaction, in
general, is weaker than the interaction among the pseudoscalar mesons.

The fact that the two-loop graph in question does not contribute in the case of the ferromagnet, may be rather surprising to readers
trained in chiral perturbation theory, where this two-loop graph indeed does contribute to the partition function \citep{GL89}. We are
dealing here with one of the novel effects that occur in the case of nonrelativistic kinematics. In fact, any of the three terms in the
leading-order effective Lagrangian (\ref{leadingLagrangian}) yields expressions with four magnon fields that appear to lead to a nonzero
contribution for the two-loop graph. However, making use of the leading-order equation for the magnon propagator -- much like using the
leading-order equation for the pion propagator in chiral perturbation theory -- these different contributions can be reduced to a single
one, proportional to $[ {\partial}_r G(x) ]_{x=0} \, [{\partial}_r G(x) ]_{x=0}$, where the time derivative and the magnetic field have
been eliminated. One therefore concludes that this two-loop graph, in the context of ferromagnets, does not contribute to the partition
function due to parity. On the other hand, the same two-loop diagram in chiral perturbation theory is proportional to $M^2 [G(x)]_{x=0}
[G(x)]_{x=0}$.

We now consider the low-temperature series for the pressure, starting with the ferromagnet. While the leading terms in the pressure
$P_{d_s}$ are of order ($T^{5/2},T^2,T^{3/2}$) in $d_s$=(3,2,1), the spin-wave interaction only emerges at ($T^5 , T^4ln T,T^{5/2}$),
\begin{eqnarray}
P_3 & = & a_0 T^{\frac{5}{2}} + a_1 T^{\frac{7}{2}} + a_2 T^{\frac{9}{2}} + {\bf a_3 T^5} + {\bf a_4 T^{\frac{11}{2}}} + {\cal O}({\bf p^{12}}) , 
\nonumber \\
\label{pressureD2Log}
P_2 & = & {\hat a}_0 T^2 + {\hat a}_1 T^3 + {\bf {\hat a}^A_2 \, T^4} + {\bf {\hat a}^B_2 \, T^4 \, ln T} + {\cal O}({\bf p^{10}}) ,
\nonumber \\
P_1 & = & {\tilde a}_0 T^{\frac{3}{2}} + {\bf {\tilde a}_1 T^{\frac{5}{2}}} + {\cal O}({\bf p^6}) .
\end{eqnarray}
Note that all interaction terms are boldfaced. Since the two-loop diagram with an insertion from ${\cal L}^2_{\mbox{\scriptsize eff, F}}$
does not contribute (diagram 6a in Fig.~\ref{figure2}), interaction terms of order ($T^4,T^3,T^2$) in $d_s$=(3,2,1) are absent. One
further notices that the higher the spatial dimension, the less important the interaction becomes: there are more terms related to
noninteracting magnons, until the interaction sets in. In particular, as Dyson showed a long time ago \cite{Dys56}, the spin-wave
interaction in the case of the three-dimensional ferromagnet only manifests itself at order $T^5$ in the pressure, i.e., far beyond the
leading Bloch term of order $T^{\frac{5}{2}}$.

Interestingly, ferromagnets in two spatial dimensions develop a logarithm, because in the limit ${\vec H} \! \to \! 0$ the cateye graph
diverges logarithmically like $T^4 ln(\mu H/T)$ \citep{Hof12b}. In chiral perturbation theory, the situation is analogous: in the chiral
limit ($m_q,M_{\pi} \to 0$), the cateye graph diverges like $T^8 ln(M_{\pi}/T)$ \citep{GL89}, such that we also have a logarithmic
contribution in the low-temperature series of the pressure,
\begin{equation}
\label{pressureQCD}
P_{\mbox{\scriptsize{QCD}}} = b_0 T^4 + {\bf b_1 T^6} + {\bf b_2 \, T^8 \, ln T} + {\cal O}({\bf p^{10}}) .
\end{equation}
But, unlike for $d_s$=2 ferromagnets, an analogous term $T^8$ (without $ln$) is absent. This is because in chiral perturbation theory all
effective ${\cal L}^4_{\mbox{\scriptsize eff}}$ constants undergo logarithmic renormalization, whereas for $d_s$=2 ferromagnets only part of
them \citep{Hof12b}.

In $d_s$=1,3, the leading terms involve half-integer powers of the temperature, while loops are further suppressed by powers of
$p \propto T^{1/2}$ ($p^3 \propto T^{3/2}$) in $d_s$=1(3). For odd spatial dimensions the pressure for ferromagnets thus exhibits both
integer and half-integer powers of $T$. In the case of ferromagnetic films, on the other hand, the leading term in the pressure is of
order $T^2$. Since here loops are suppressed by one power of $T$, the series does not involve any half-integer temperature powers.
However, as we have pointed out before, logarithmic contributions emerge. In the limit $H, m_q \! \to \! 0$, the coefficients
$a_i,{\hat a}_i,{\tilde a}_i,b_i$ formally become $T$-independent, while for $H, m_q \neq 0$ they are complicated functions of the ratios
$\mu H/T$ and $M_{\pi}/T$ \citep{GL89,Hof11,Hof12b,Hof13}. Note however, as we discuss below, that the limit $H \! \to \! 0$ is
problematic in $d_s \leq 2$.

Finally, the low-temperature series of the magnetization of ferromagnets,
\begin{equation}
{\overline \Sigma_{d_s}} = \frac{\Sigma(T,H)}{\Sigma} ,
\end{equation}
in three, two and one spatial dimensions, amount to
\begin{eqnarray}
\label{magnetizations}
{\overline \Sigma_3} & = & 1 - {\alpha}_0 T^{\frac{3}{2}} - {\alpha}_1 T^{\frac{5}{2}} - {\alpha}_2 T^{\frac{7}{2}} - {\bf {\alpha}_3 T^4}
- {\bf {\alpha}_4 T^{\frac{9}{2}}} + {\cal O}({\bf T^5}) , \nonumber \\
{\overline \Sigma_2} & = & 1 - {\hat \alpha}_0 T - {\hat \alpha}_1 T^2 - {\bf {\hat \alpha}^A_2 \, T^3} - {\bf {\hat \alpha}^B_2
\, T^3 \, ln T} + {\cal O}({\bf T^4}) , \nonumber \\
{\overline \Sigma_1} & = & 1 - {\tilde \alpha}_0 T^{\frac{1}{2}} - {\bf {\tilde \alpha}_1 T^{\frac{3}{2}}} + {\cal O}({\bf T^2}) .
\end{eqnarray}
The low-temperature expansion for the analogous quantity in QCD, the quark condensate
\begin{equation}
\frac{\langle \bar{q} q \rangle(T,m_q)}{\langle 0 | \bar{q} q | 0 \rangle} ,
\end{equation}
reads
\begin{equation}
\label{quarkCondensate}
1 - {\beta}_0 T^2 - {\bf {\beta}_1 T^4} - {\bf {\beta}_2 T^6 \, ln T} + {\cal O}({\bf T^8}) .
\end{equation}
Again, for nonzero magnetic field and nonzero quark masses, the coefficients ${\alpha}_i,{\hat \alpha}_i,{\tilde \alpha}_i$ and
${\beta}_i$ depend on the ratios $\mu H/T$ and $M_{\pi}/T$ in a nontrivial way. In the chiral limit ($m_q,M_{\pi} \to 0$), the leading
coefficient ${\beta}_0$ in the quark condensate reduces to $1/8 {\cal F}^2$. The leading coefficient in the spontaneous magnetization of
the $d_s$=3 ferromagnet becomes $\alpha_0 = \zeta(\mbox{$ \frac{3}{2}$}) /8 {\pi}^{\frac{3}{2}} \Sigma {\gamma}^{\frac{3}{2}}$. The
temperature scale in chiral perturbation theory is thus given by 
$\Lambda^T_{\mbox{\scriptsize{QCD}}} = \sqrt{8} {\cal F} \approx 250 \mbox{\normalsize MeV}$, which roughly corresponds to the temperature
where the chiral phase transition takes place \citep{GL89}. In the case of the $d_s$=3 ferromagnet, the temperature scale is
$\Lambda^T_{\mbox{\scriptsize{F}}} = {\alpha}^{-2/3}_0 \approx 10 \mbox{\normalsize meV}$, which is roughly of the order of the Curie
temperature where the spontaneous magnetization becomes zero. Although these scales differ in more than ten orders of magnitude, the
effective theory captures both systems from a universal perspective based on symmetry. In particular, for temperatures small compared to
the corresponding scales $\Lambda^T_{\mbox{\scriptsize{QCD}}}$ and $\Lambda^T_{\mbox{\scriptsize{F}}}$, the series presented in this review are
perfectly valid.

It is important to point out that in one and two spatial dimensions, the leading coefficients ${\hat \alpha}_0$ and ${\tilde \alpha}_0$ in
the magnetization become divergent in the limit ${\vec H} \! \to \! 0$. This is related to the Mermin-Wagner theorem \citep{MW68}, stating
that spontaneous symmetry breaking in $d_s \le 2$ cannot occur at finite temperature in the Heisenberg model. According to
Eqs.~(\ref{magnetizations}), the magnetization $\Sigma(T,H)$, in any dimension $d_s$=1,2,3, tends to zero at a "critical" temperature
$T_c$, where the spontaneously broken symmetry is restored. While $T_c$ in $d_s$=3 tends to a finite value in the limit
${\vec H} \! \to \! 0$ (much like for the quark condensate, $T_c$ tends to a finite value in the chiral limit), the "critical" temperatures in
$d_s$=1,2 tend to zero if the magnetic field is switched off. Spontaneous symmetry breaking never occurs here at finite temperature --
that is how the effective theory "knows" about the Mermin-Wagner theorem.

In addition, in $d_s \leq 2$, an energy gap is generated nonperturbatively at finite temperature \citep{Kop89,KC89}. Strictly speaking
this also implies that the limit ${\vec H} \! \to \! 0$ cannot be taken in $d_s \le 2$, neither in the pressure nor in the magnetization,
because we would then leave the domain of validity where the effective expansion applies \citep{Hof12a,Hof12b,Hof13}. Still, in two
spatial dimensions, where the the nonperturbatively generated correlation length $\xi_{np}$ is exponentially large
($\xi_{np} \propto e^{1/T}$), this effect is tiny in the pressure, and does not numerically affect the corresponding low-temperature
series. In the context of ferromagnetic spin chains, however, where the nonperturbatively generated correlation length is proportional to
the inverse temperature ($\xi_{np} \propto 1/T$), it would be completely inconsistent to switch off the magnetic field even in the
low-temperature series of the pressure \citep{Hof12a,Hof12b,Hof13}. 

It is important to point out that these subtleties in one and two spatial dimensions only emerge at finite temperature. In particular, at
$T = 0$ the limit ${\vec H} \! \to \! 0$ is well defined in $d_s \le 2$. Much like in chiral perturbation theory or in ferromagnetic
systems  in $d_s=3$, no divergent behavior of the observables occurs. In other words, the "chiral limit" ${\vec H} \! \to \! 0$ is not
problematic for ferromagnets living in one or two spatial dimensions at zero temperature. However, if the temperature is finite, then the
magnetic field cannot be totally switched off in $d_s \le 2$, because one would then leave the domain of validity of the effective
expansion. As discussed in detail in Refs.~\citep{Hof12a,Hof12b,Hof13}, this parameter regime where the effective expansion breaks down,
is tiny. Again, this observation is related to the Mermin-Wagner theorem and the nonperturbatively generated energy gap.

\section{Conclusions}

The systematic effective Lagrangian method is a very useful tool also in the nonrelativistic domain where it even works in one spatial
dimension. The method is appealing due to its universal character, interconnecting different branches of physics, such as particle and
condensed matter physics. Chiral logarithms showing up in the renormalization of next-to-leading order effective constants in chiral
perturbation theory and $d_s$=2 ferromagnets, e.g., are due to analogous ultraviolet divergences. In general, the structure of the
low-temperature series -- both in relativistic and nonrelativistic effective field theory -- is an immediate consequence of the
spontaneously broken symmetry. Regarding ferromagnets, the effective method proves more powerful than conventional condensed matter
approaches, where analogous three-loop calculations are either missing or are erroneous.

\end{document}